\documentclass{ws-p8-50x6-00}

\def\CPC{{\em Comp. Phys. Comm.} }
\def\EPJC{{\em Eur. Phys. J.} C}
\def\IJMPA{{\em Int. J. Mod. Phys.} A}

\def\NPB{{\em Nucl. Phys.} B}
\def\PLB{{\em Phys. Lett.} B}

\def\PRp{{\em Phys. Rep.} } 

\def\ZPC{{\em Z. Phys.} C}

\usepackage{amsmath}

\begin{document}

\title{QCD DESCRIPTION OF ANGULAR CORRELATIONS}

\author{WOLFGANG OCHS}

\address{Max-Planck-Institut f\"ur Physik (Werner Heisenberg Institut)\\
F\"ohringer Ring 6, D80805 M\"unchen, Germany\\
E-mail: wwo@mppmu.mpg.de}


\maketitle

\abstracts{
We review the predictions on angular correlations and their recent
experimental tests at LEP and HERA. Power behaviour of correlation functions 
appear for large angles and reflect
the underlying fractal structure in jet evolution. Asymptotic scaling laws
work rather well for the low momentum particles.
For angular correlations the predictions work at least qualitatively,
sometimes quantitatively, at LEP and the higher HERA
energies. Some limitations of the approach and possible improvements 
are discussed.}

\section{Introduction}

There has been much interest in the last decade in the 
study of correlation functions and their possible power dependence
on resolution scale.
Such behaviour is expected, in particular, from the fractal structure of
the selfsimilar
parton cascade.\cite{lufractal} Specific predictions, mainly within the Double
Logarithmic approximation (DLA) of
 pQCD, have been derived for
angular correlations and multiplicity 
moments.\cite{ow,dd,bp}
%
In recent years these predictions have been tested in 
various experimental works by the
 L3\cite{L312}, DELPHI\cite{delphi12} and ZEUS\cite{zeus12}
Collaborations, see also talk at
this conference\cite{leszek}, so there is now
a good time for a critical assessment on what has been learned from these
studies.

A major aim is the test of ``Local Parton Hadron Duality''
(LPHD$\,$\cite{adkt1}) which suggests comparing directly the multi-hadron
with the multi-parton final state without the interface of a ``hadronization
model.'' 
This approach has 
been proposed originally for inclusive spectra 
but has been applied subsequently to many other problems 
(recent reviews\cite{ko}). Here we adress its application to
multi-particle correlations beyond single inclusive phenomena.

\section{Correlations in Full and Restricted Angular Range}
Particle multiplicity distributions 
can be characterized by their factorial
moments $F_q=\langle n(n-1)\ldots(n-q+1)\rangle /\langle n\rangle^q$
which relate to integrals of the  densities 
$(dn/dp_1\ldots dp_q)/\langle n\rangle^q$. 
Perturbative QCD predictions for the moments 
in full phase space (event or jet) improve strongly with increasing
accuracy of the logarithmic approximations (see review\cite{dg})
and are entirely satisfactory for the
exact numerical
solution of the pertaining evolution equations.\cite{sl} The predictions
depend on the QCD scale $\Lambda$ and one non-perturbative parameter
($k_\perp$ cut-off $Q_0$); 
they can be determined from a fit to
the global event multiplicities. 

For limited phase space
different observables have been considered:\\ 
\noindent {\it Distributions in relative angle $\vartheta_{12}$ between
two particles}, both inside the forward cone of a jet 
with half opening angle $\Theta$ and momentum $P$. The distribution is
normalized either by the full multiplicity
inside this cone 
$\hat r(\vartheta_{12}) = (dn/\vartheta_{12})/{\overline n(\Theta)}$
or by the corresponding distribution of uncorrelated particles
$r(\vartheta_{12}) = {(dn/\vartheta_{12})}/{(dn/\vartheta_{12})_{uncorr}}$.\\
\noindent {\it Multiplicity moments $F_q$}, from particles
inside an angular ring of size $2\delta$
at polar angle $\Theta$ to the jet axis or from inside a cone of half angle
$\delta$ again at polar angle $\Theta$; they correspond to
dimensions $D=1$ and $D=2$ respectively.

In the theory with fixed coupling $\alpha_s$ all these observables
(generically denoted by $h_q$) 
show a universal power behaviour ($\vartheta_{12}\to \delta$)
\begin{gather}
  h_q(\delta,\Theta,P) \sim  \left(\Theta/\delta\right)^{\varphi_q}
\qquad \qquad\hfill
\label{power}\\
{\rm for}\ F_q:\quad  \varphi_q  = (q-1)D -(q-q^{-1}) \gamma_0;\qquad 
{\rm for}\ \hat r:\quad \varphi_2=-3\gamma_0/2
\label{cordefs}
\end{gather}
corresponding to the selfsimilar structure of the parton cascade.
The power is given in terms of the ``QCD anomalous dimension''
$\gamma_0=\sqrt{6\alpha_s/\pi}$.    

For running coupling this result is retained only at the large angles 
$\delta\sim{\cal O}(\Theta)$ where the coupling varies as
$\alpha_s(P\Theta/\Lambda)$. However,
with decreasing angles $\delta$ there is 
an increasing deviation from the power law because of the logarithmic growth 
of the coupling for small angular scales $k_\perp \sim P\delta$.
The authors\cite{ow,dd,bp} obtain slightly different approximations,
for example,\cite{ow}
\begin{equation}         
h_q(\delta,\Theta,P)\sim \exp(2q\gamma_0(P\Theta/\Lambda)
    \omega(\epsilon,q)),\qquad \varepsilon=
       \frac{\ln(\Theta/\delta)}{\ln(P\Theta/\Lambda)}
\label{runalph}
\end{equation}           
where $\omega(\epsilon,q)$ is the solution of an algebraic equation and
reproduces the limit (\ref{power}) for large angles.

\section{Discussion of Experimental Results}
\noindent {\it Angular dependence (on $\delta, \vartheta_{12}$)}.
In all measurements at LEP and HERA the data behave qualitatively  
as predicted: for large angles (small $\varepsilon$) there is an approximate
power behaviour and a deviation at small angles according to the prediction.
The deviation between predictions and experiment varies between 10-20\%
for $\ln F_q,\ q\geq 3$ and a bit more for $F_2$, the differential correlations
$r,\ \hat r$ at the higher HERA energies ($Q^2>2000$ GeV$^2$) 
agree rather well with the prediction but deviate considerably at low
$Q^2$, according to what is  expected for an asymptotic prediction.\\
\noindent {\it Initial slope  $\varphi_q$}. 
The powers $\varphi_q$ in (\ref{power}) have
been extracted by DELPHI\cite{delphi12} 
from the  $F_q$ moments at small $\varepsilon$ for $q\leq 5$
and $D=1,2$, see Fig. 1. The overall variation of the slopes
by a factor $\sim$20 is roughly met by the prediction (\ref{cordefs})
and the deviations grow up to $\sim$30\% at high $q$.
Furthermore it is found,\cite{delphi12} that the initial slope decreases
with increasing  $\Theta$ as expected from the running coupling
 $\alpha_s(P\Theta/\Lambda)$, best results are obtained for $\Lambda=0.04$
GeV which is a bit small but still acceptable for a DLA calculation,
also $n_f=3$ is taken.\\
\begin{figure*}[t]
\begin{center}
\mbox{\epsfig{file=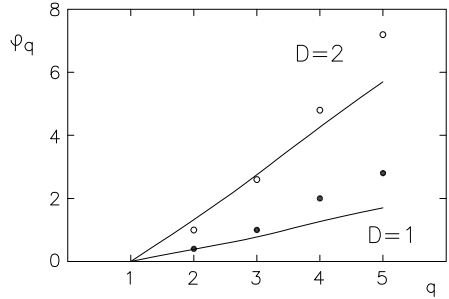,%
bbllx=2.2cm,bblly=13.5cm,bburx=16.5cm,bbury=23.5cm,%
width=6.cm}}  
          \end{center}
\vspace{-0.3cm}
\caption{
Initial slope $\varphi_q$ of factorial moments 
from DELPHI\protect\cite{delphi12} in comparison with asymptotic predictions
(\ref{power},\ref{cordefs}) for angular  ring ($D=1$) and cone ($D=2$) at order
$q$, curves connect results at integer $q$ (parameters
$n_f=3,\ \Lambda=0.04$~GeV).} 
\label{phiq}
\end{figure*}
\noindent {\it Dependence on approximation scheme}. 
The approximations (DLA)
are justified only at asymptotic energies 
and could be responsible for the partial disagreement 
with experiment.
The ZEUS collaboration also compared their results on $F_q$ with the
predictions of a parton level MC (ARIADNE\cite{ariadne}) 
using a small cut-off parameter $Q_0$ as in the analytic 
calculations. Then agreement with
data improves considerably, with results
comparable, for example, to the LEPTO MC with full hadronization.

\section{Asymptotic scaling laws}
The DLA results correspond to the high energy asymptotics. These results can
often be written in scaling form. A well known example is the KNO scaling,
which says, that the rescaled  probability $\langle n\rangle P_n$ approaches an energy
independent (scaling) limit as function of the rescaled multiplicity
$n/\langle n\rangle $. In a similar way one finds an asymptotic limit for 
the energy spectra in appropriate rescaled variables
(``$\zeta$-scaling''\cite{dfk2}). The data follow this scaling
prediction only for the low momentum particles, a phenomenon understood as a
consequence of soft gluon coherence.\cite{ko}
Typically, 
 at all available primary energies
the deviations from the scaling limit  increase with
momentum (they
can be taken into account by the higher order (MLLA) corrections).\cite{lo}

For angular correlations Eq. 
(\ref{runalph}) can be rewritten in scaling form
\begin{equation}
\ln h_q(\delta,\Theta,P)/\gamma_0(\Theta,P) = f(\epsilon,q)
\label{epsscal}
\end{equation}
with known $f$,
 so the rescaled correlations in the variable $\varepsilon$ 
become independent of both the jet energy $P$ and
the large angle $\Theta$ (``$\varepsilon$-scaling''\cite{ow}).

Remarkably, the correlation functions $r,\ \hat r$ approach the 
predicted scaling limit within the HERA $Q^2$ range,
furthermore, they show the 
independence of angle $\Theta$ in the range $\Theta=30^\circ\ldots 90^\circ$, also at LEP.
We relate this good agreement already at present energies
to the good scaling properties 
of low momentum  particles which dominate the
correlation measurements. 

\section{Limitations and Improvements}
A problem has recently been encountered with
moments for particles in the cylinder $p_\perp<p_\perp^{cut}$.
 A recent measurement by
ZEUS\cite{zeus12,leszek} has revealed a rise of these moments for small
cut-off $p_\perp^{cut}$ contrary to the prediction\cite{low} of a decrease
 $F_q\to 1$ for  $p_\perp^{cut}\to 0$ (or $p_\perp^{cut}\to Q_0$ in the 
calculation) corresponding to a Poissonian  multiplicity distribution;
this prediction was also confirmed by a parton MC calculation.\cite{low}

We take this observation as showing a limitation of the duality approach.
For sure, there  cannot be a perfect matching
between parton and hadron final states, as already obvious from the
existence of hadronic resonances. The correlations of 
particles at low $p_\perp$ seem to involve additional non-perturbative effects.
Perhaps one can find a better treatment of particles near the kinematic
cut-off $p_\perp \sim Q_0$ 
as it was possible for single inclusive spectra in this limit
where similar problems occur. 
Beyond the original discussion,\cite{low} we would expect
the perturbative prediction of the  Poissonian limit to apply for jets
of not too low virtuality $y_{cut}\gg (Q_0/Q)^2$ where 
the dependence on $Q_0$ disappears.

On the other hand, the perturbative DLA predictions for
correlations in angle are in overall agreement with measurements.
A considerable improvement of the
calculations is not obtained by including MLLA corrections.\cite{dd}  
We would expect a major improvement in accuracy,
especially for the large angle behaviour,
if full matrix element calculations were carried out as it has been done 
for azimuthal angle correlations
with good success.\cite{dmo} An easier path towards higher
accuracy is available by running a parton MC (ARIADNE) 
with non-standard parameters $Q_0\sim \Lambda$ 
as discussed previously.\cite{low} 

\section{Summary}
The dual connection between
parton and hadron final states is not restricted to single particle
distributions but extends to multi-particle correlations; 
care has to be taken with
particles near the kinematic border at $p_\perp\sim Q_0$.  
The simple asymptotic formulae from DLA for angular correlations 
describe all data 
 at least at the qualitative level.
Near quantitative results follow for global observables or from
MC calculations.
The fixed coupling results correspond to 
a selfsimilar cascade and fractal structure. The realistic QCD cascade with
running coupling corresponds to this scaling picture approximately but shows 
characteristic deviations in angular distributions and with primary energy.

\end{document}